\documentclass[preprint,review,12pt]{elsarticle}

\usepackage{amssymb}

\usepackage{amsmath,amssymb,amsfonts}
\usepackage{algorithmic}
\usepackage{graphicx}
\usepackage{ulem}
\usepackage{soul}
\usepackage{textcomp}
\usepackage{xcolor}
\usepackage{float}
\usepackage{overpic}
\usepackage{multirow}
\usepackage{subfigure}
\usepackage{booktabs}

\usepackage[colorlinks,
linkcolor=red,
anchorcolor=red,
citecolor=green]{hyperref}
\def\BibTeX{{\rm B\kern-.05em{\sc i\kern-.025em b}\kern-.08em
		T\kern-.1667em\lower.7ex\hbox{E}\kern-.125emX}}

\journal{Expert Systems with Applications}

\begin{document}

\title{Retinal Vessel Segmentation with Deep Graph and Capsule Reasoning}

\author[1,3]{Xinxu Wei}
\ead{xxwei523@gmail.com}

\author[3]{Xi Lin}
\ead{xi.lin@std.uestc.edu.cn}

\author[3]{Shixuan Zhao}
\ead{phanzsx@gmail.com}

\author[2]{Haiyun Liu}
\ead{1335914484@qq.com}


\author[1,3]{Yongjie Li*}
\ead{liyj@uestc.edu.cn}

\tnotetext[t1]{This work was supported by Huzhou Science and Technology Program (\#2023GZ13).}

\cortext[cor1]{Corresponding author}

\address[1]{Yangtze Delta Region Institute (Huzhou), University of Electronic Science and Technology of China, Chengdu, China}



\address[2]{Department of Computer Science and Engineering, University of South Florida, Tampa, FL, USA}

\address[3]{School of Life Science and Technology, University of Electronic Science and Technology of China, Chengdu, China}

\begin{abstract}
Effective retinal vessel segmentation requires a sophisticated integration of global contextual awareness and local vessel continuity. To address this challenge, we propose the Graph Capsule Convolution Network (GCC-UNet), which merges capsule convolutions with CNNs to capture both local and global features. The Graph Capsule Convolution operator is specifically designed to enhance the representation of global context, while the Selective Graph Attention Fusion module ensures seamless integration of local and global information.
To further improve vessel continuity, we introduce the Bottleneck Graph Attention module, which incorporates Channel-wise and Spatial Graph Attention mechanisms. The Multi-Scale Graph Fusion module adeptly combines features from various scales. Our approach has been rigorously validated through experiments on widely used public datasets, with ablation studies confirming the efficacy of each component. Comparative results highlight GCC-UNet's superior performance over existing methods, setting a new benchmark in retinal vessel segmentation. Notably, this work represents the first integration of vanilla, graph, and capsule convolutional techniques in the domain of medical image segmentation.
\end{abstract}


\maketitle

\section{Introduction}
\label{sec:introduction}
Retinal vessel segmentation is a key step in diagnosing retinal diseases like diabetic retinopathy and glaucoma, as changes in the vascular structure offer important diagnostic insights \cite{li2021applications}. However, manually segmenting vessels is often labor-intensive and prone to mistakes, especially when addressing thin, low-contrast vessels against the intricate background of the fundus. Therefore, developing automated and accurate segmentation algorithms is critical for improving clinical workflow.

The segmentation task is challenging due to the intricate structure of retinal vessels, which often blend into the background or are obscured by lesions. Thin vessels, especially capillaries, are hard to detect and frequently mislabeled due to their similar characteristics to the surrounding tissues \cite{akbar2019automated, lin2024vascular}. Although traditional methods \cite{soares2006retinal}\cite{chaudhuri1989detection} and machine learning-based techniques have shown some success \cite{orlando2014learning}, they often rely on pre-defined features and struggle with the fine details necessary for accurate segmentation.

Recently, deep learning techniques have become dominant in the field, achieving cutting-edge performance in medical image segmentation \cite{wang2020csu}\cite{feng2020ccnet}. Despite these advances, two key challenges remain: capturing comprehensive global context and ensuring vessel continuity, particularly for the smallest vessels. Methods such as dilated convolutions \cite{yu2015multi}, attention mechanisms \cite{vaswani2017attention}, and non-local operations \cite{wang2018non} have been proposed to address these issues. However, these methods often fail to fully model the part-to-whole relationships essential for context in retinal images, while existing loss functions and attention modules struggle to maintain vessel continuity amidst noise from other tissues or lesions.
To tackle these challenges, we propose a Graph Capsule Convolution UNet (GCC-UNet), which introduces capsule convolutions for part-to-whole modeling and graph reasoning to enhance vessel continuity. This framework uniquely combines vanilla convolution, capsule convolution, and graph convolution, leveraging their complementary strengths to achieve more robust segmentation.

The main contributions of this work are:

\begin{itemize} 
\item We present the GCC-UNet, which captures both local features and global contextual information for retinal vessel segmentation. 
\item We introduce the Graph Capsule Convolution (GC-Conv) operator, integrating graph reasoning into capsule networks to improve the representation of global vessel structures. \item We develop the Selective Graph Attention Fusion (SGAF) module, designed to effectively merge global and local context features. 
\item We propose a Bottleneck Graph Attention (BGA) module, which improves vessel continuity through Channel-wise and Spatial Graph Attention mechanisms. 
\item We design a Multi-Scale Graph Fusion (MSGF) module that combines features at different scales to enhance segmentation performance. 
\item Extensive experiments on various datasets demonstrate that our approach surpasses existing methods, setting a new benchmark in retinal vessel segmentation. \end{itemize}


\begin{figure*}[htbp]
	\centering
	\includegraphics[width=\linewidth]{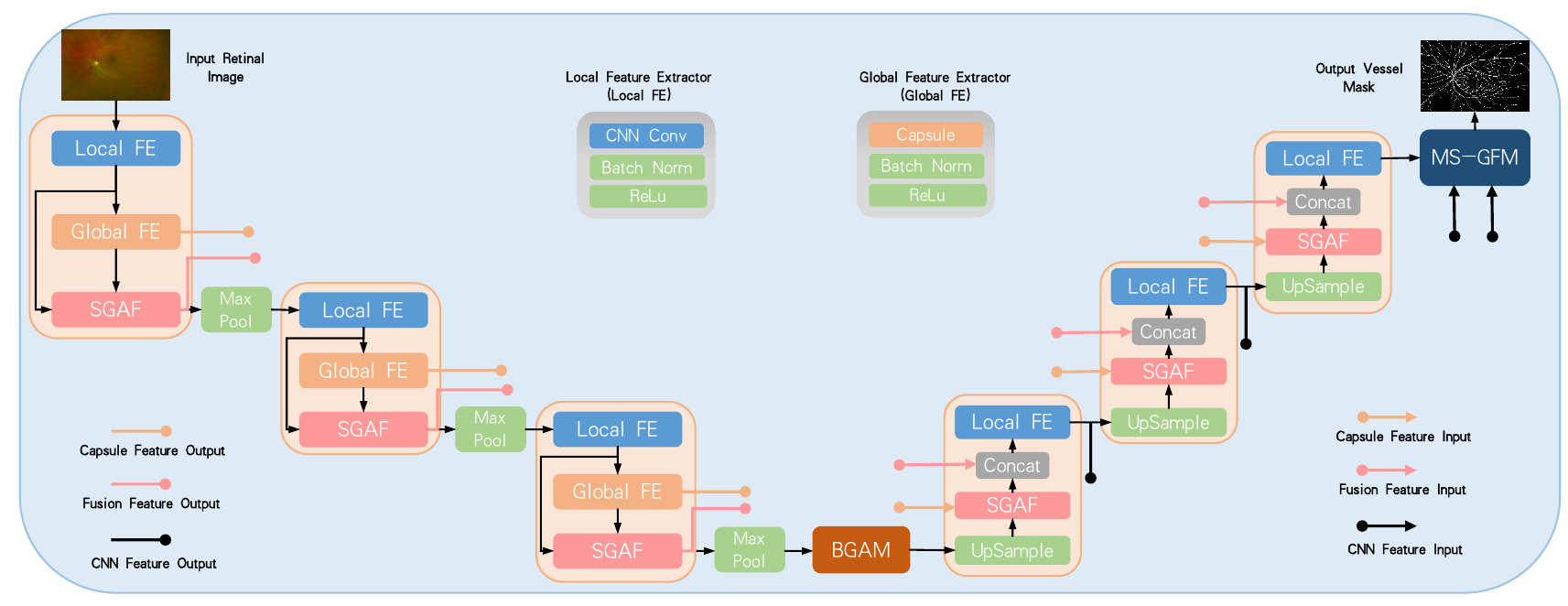}
	\caption{The network architecture of the proposed GCC-UNet.}
	\label{model}
\end{figure*}

\section{Related Works} \label{sec
works}


Traditional approaches utilize a variety of image processing techniques, such as filtering \cite{chaudhuri1989detection, zhang2010retinal} and handcrafted feature extraction \cite{soares2006retinal, zhang2010retinal}, to distinguish retinal vessels from the background. Numerous studies have explored vessel attributes like orientation \cite{yin2015vessel}, edge detection \cite{xie2015holistically}, line patterns \cite{ricci2007retinal, nguyen2013effective}, vessel width \cite{khan2022width}, and structural topology \cite{estrada2015retinal}, contributing to improved segmentation. Furthermore, machine learning-based approaches \cite{orlando2016discriminatively, fraz2012ensemble, estrada2015retinal}, including classifiers such as Support Vector Machines (SVMs) \cite{ricci2007retinal}, have demonstrated efficacy in vessel segmentation tasks.

In recent years, deep learning techniques \cite{simonyan2014very, he2016deep, wei2022deep} have gained prominence due to their robust feature extraction capabilities. For instance, DeepVessel \cite{fu2016deepvessel} adopts a HED-like architecture combined with Conditional Random Fields (CRFs) for vessel detection. Other deep learning approaches, such as DRIU \cite{maninis2016deep}, VGN \cite{shin2019deep}, V-GAN \cite{son2017retinal}, BTS-DSN \cite{guo2019bts}, SWT-FCN \cite{oliveira2018retinal}, DeepDyn \cite{khanal2020dynamic}, and DRIS-GP \cite{cherukuri2019deep}, have achieved noteworthy results. UNet \cite{ronneberger2015u}, a well-established model in medical image segmentation, has inspired various UNet-based models for retinal vessel segmentation, including Attention UNet \cite{oktay2018attention}, Dense UNet \cite{li2018h}, Deformable UNet \cite{jin2019dunet}, SA-UNet \cite{guo2021sa}, JL-UNet \cite{yan2018joint}, CC-Net \cite{feng2020ccnet}, CTF-Net \cite{wang2020ctf}, CSU-Net \cite{wang2020csu}, OCE-Net \cite{wei2023orientation} and RCAR-UNet \cite{ding2024rcar}.

Despite these advancements, many methods struggle to fully capture the intricate relationships between vessel characteristics and ensure vessel continuity, especially when external noise is present.

\subsection{Graph Neural Networks} 
Graph neural networks (GNNs), particularly graph convolutional networks (GCNs) \cite{zhang2019graph, kipf2016semi}, have garnered considerable attention across diverse fields such as computer vision \cite{zhang2019dual, ye2020attention}. Initially proposed by Kipf \cite{kipf2016semi} for the classification of non-Euclidean data, GCNs have since been employed in tasks including image recognition \cite{ye2020attention}, segmentation \cite{zhang2019dual}, and medical image analysis \cite{meng2021bi}. Despite the significant advancements in GNN research \cite{velivckovic2017graph, you2019position}, their application in vessel segmentation remains relatively underexplored \cite{shin2019deep}.

Our research addresses this gap by incorporating a GCN module specifically designed for retinal vessel segmentation, leveraging the geometric modeling capabilities of GCNs. By capturing the structural features of vessels, we aim to enhance vessel continuity and minimize interference from surrounding tissues.

\subsection{Capsule Neural Networks} 
Capsule networks, which are specifically designed to capture spatial relationships between objects, excel in distinguishing multiple overlapping entities within an image. Hinton \cite{sabour2017dynamic} introduced the concept of capsules to address the limitations of CNNs, particularly their restricted ability to capture global context due to limited receptive fields and their lack of equivariance. Sabour’s \cite{sabour2017dynamic} capsule network architecture employs dynamic routing between capsule units to represent part-whole relationships, thereby enhancing equivariance. Since this introduction, several enhancements such as Attentive Capsule Networks \cite{choi2019attention}, Graph-Capsule Networks \cite{xinyi2019capsule, verma2018graph}, and DeformCaps \cite{lalonde2021deformable}, along with innovations in routing mechanisms (e.g., EM-Routing \cite{hinton2018matrix} and Self-Attention Routing \cite{mazzia2021efficient}), have broadened the applicability of capsule networks in tasks such as object detection, image classification, and medical image segmentation \cite{lalonde2021capsules, du2019novel}.

However, most capsule-based methods focus on optimizing routing algorithms rather than addressing the relationships between capsule elements. Inspired by retinal vessel characteristics, we introduce capsule networks to retinal vessel segmentation, employing their global context modeling capabilities. By incorporating graph reasoning into capsule networks, our approach captures interactions between capsule elements, enhancing vessel continuity while maintaining context awareness.


\section{Methodology}

\label{sec:methods}

\subsection{Overall Architecture}

The GCC-UNet architecture, illustrated in Figure \ref{model}, builds upon the U-Net \cite{ronneberger2015u} as its foundational framework. In the downsampling process, a Local Feature Extractor (Local FE), a Global Feature Extractor (Global FE), and a Selective Graph Attention Fusion (SGAF) module are introduced to combine local features captured by a conventional CNN with global context features derived from a Capsule Neural Network. To achieve global feature extraction, we propose a Graph Capsule Convolution (GC Conv) operator, which replaces the standard capsule convolution operator. Additionally, a Bottleneck Graph Attention (BGA) module is integrated into the bottleneck to enhance vessel continuity by modeling the connectivity of vessel nodes along the graph. In the upsampling phase, the global context features are directly passed to the upsampling layer, minimizing computational overhead. The SGAF then fuses the global features with the upsampled local features. Lastly, a Multi-Scale Graph Fusion (MSGF) module is employed to integrate features across different stages of the U-Net.

\subsection{Graph Capsule Convolution}

\begin{figure*}[htbp]
	\centering
	\includegraphics[width=14cm]{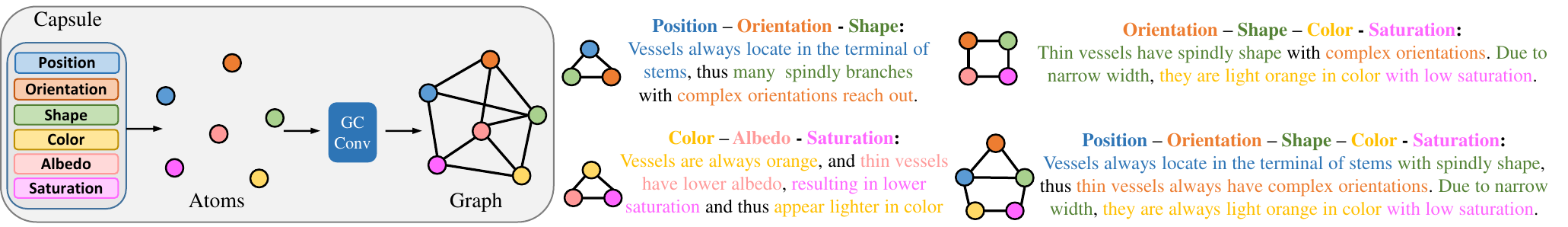}
	\caption{The schematic illustrates the relationships among atoms within capsules. By incorporating graph structures into capsules, we model the part-to-whole relationships among various characteristics of atoms, distinguishing between vessel components and the fundus background.}
	\label{relation}
\end{figure*}

\begin{figure}[htbp]
	\centering
	\includegraphics[width=14cm]{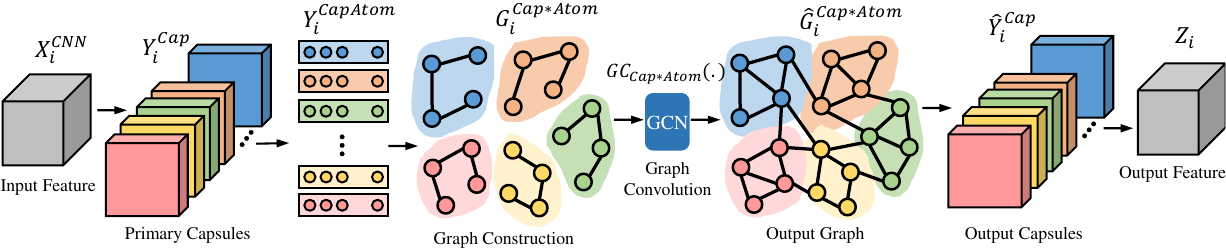}
	\caption{The proposed Graph Capsule Convolution (GC Conv).}
	\label{GCConv}
\end{figure}

Unlike traditional CNNs that utilize scalar elements, capsule networks (CapsNets) employ vectors as their fundamental components. Each capsule contains a vector that captures various intrinsic characteristics of an object, such as its pose (position, size, orientation, shape), deformation, color, and saturation. The length of the vector represents the probability of the object's presence in the image. While CapsNets excel in capturing detailed local features, they struggle with modeling translation invariance and part-to-whole relationships compared to CNNs.

To improve global feature extraction for retinal vessels, we enhance the standard capsule convolution \cite{sabour2017dynamic} by integrating graph representation learning to model the relationships among capsules. We propose a novel Graph Capsule Convolution (GC Conv).

As illustrated in Fig. \ref{GCConv}, the input features extracted by a conventional CNN are transformed into primary capsules, representing low-level entities. Dynamic routing \cite{sabour2017dynamic} then directs these low-level capsules to high-level ones, capturing part-to-whole relationships. This dynamic routing operates as a transfer matrix with attention weights, emphasizing important capsules and vectors while disregarding less relevant ones. However, the original dynamic routing \cite{sabour2017dynamic} does not account for correlations among different capsules and atoms within capsules. To address this, we incorporate graph reasoning into the dynamic routing process, as shown in Fig. \ref{GCConv}, to better model these correlations. This approach enables us to effectively capture relationships among channels, capsules, and atoms.

As shown in Fig. \ref{GCConv}, the input CNN feature $X^{CNN}_{i}$ with a shape of $[B, C, H, W]$ is transformed into the primary capsules $Y^{Cap}_{i}$, which have a shape of $[B, H, W, k^{2}, C, L, V]$, where $B$, $H$, $W$, $C$, $L$, and $V$ represent the numbers of batch sizes, height, width, channels, capsules, and atoms in each capsule, respectively. 
\begin{equation}
	X^{CNN}_{i} \quad \xrightarrow{Transform} \quad Y^{Cap}_{i}
\end{equation}
And the channel dimension of $Y^{Cap}_{i}$ is split from the features to obtain independent features $Y^{Channel}_{i}$ with a shape of $[B, H, W, k^{2}, C, 1, 1]$, which are independent of the dimensions of capsules and atoms. 
Similarly, the dimensions of capsules and atoms are split to obtain independent features $Y^{CapAtom}_{i}$ with a shape of $[B, H, W, k^{2}, 1, L, V]$. By multiplying the channels of capsules and atoms dimensions, we obtain the feature of $Y^{Cap*Atom}_{i}$ with a shape of $[B, H, W, k^{2}, 1, L*V]$.
\begin{equation}
	Y^{Cap}_{i} \quad \xrightarrow[Channel-wise]{Split} \quad Y^{Channel}_{i}, \quad Y^{CapAtom}_{i}
\end{equation}
We then use average pooling to remove the $H$, $W$, and $K^{2}$ dimensions and construct a graph $G^{Channel}_{i}$ along the channel dimension $C$ for $Y^{Channel}{i}$. 
\begin{equation}
	\begin{aligned}
		&Y^{Channel}_{i} \quad \xrightarrow[Channel-wise]{Graph Construction} \quad G^{Channel}_{i} \\
		&Y^{Cap*Atom}_{i} \quad \xrightarrow[Channel-wise]{Graph Construction} \quad G^{Cap*Atom}_{i}
	\end{aligned}
\end{equation}
Similarly, we construct a graph $G^{Cap*Atom}_{i}$ along the $L*V$ dimension for $Y^{Cap*Atom}_{i}$. 
We apply a graph convolution $GC_{Channel}(.)$ on $G^{Channel}_{i}$ to obtain the output graph feature $\widehat{G}^{Channel}_{i}$. We also apply a graph convolution $GC_{Cap*Atom}(.)$ on $G^{Cap*Atom}_{i}$ to obtain the output graph feature $\widehat{G}^{Cap*Atom}_{i}$.
\begin{equation}
	\begin{aligned}
		&\widehat{G}^{Channel}_{i} = GC_{Channel}(G^{Channel}_{i}) \\
		&\widehat{G}^{Cap*Atom}_{i} = GC_{Cap*Atom}(G^{Cap*Atom}_{i})
	\end{aligned}
\end{equation}
Finally, we integrate $\widehat{G}^{Channel}_{i}$ and $\widehat{G}^{Cap*Atom}_{i}$ using addition and expansion operators, and transfer them into capsule features $\widehat{Y}^{Cap}_{i}$ to obtain the output feature $Z_{i}$.

\subsection{Selective Graph Attention Fusion Module}

Incorporating global context is essential for models to handle variations in scale, orientation, and partial occlusions of fundus vessels. However, capsule neural networks (CapsNets) face challenges in learning crucial local features. To address this, a promising approach is to combine capsule convolution with traditional CNN models, enabling the model to capture both local and global features effectively.

To achieve optimal fusion performance, we propose a novel Selective Graph Attention Fusion (SGAF) module. This module leverages the graph structure to model the relationships within channels of both local and global features, while also learning the correlations between these features.

\begin{figure}[htbp]
	\centering
	\includegraphics[width=11cm]{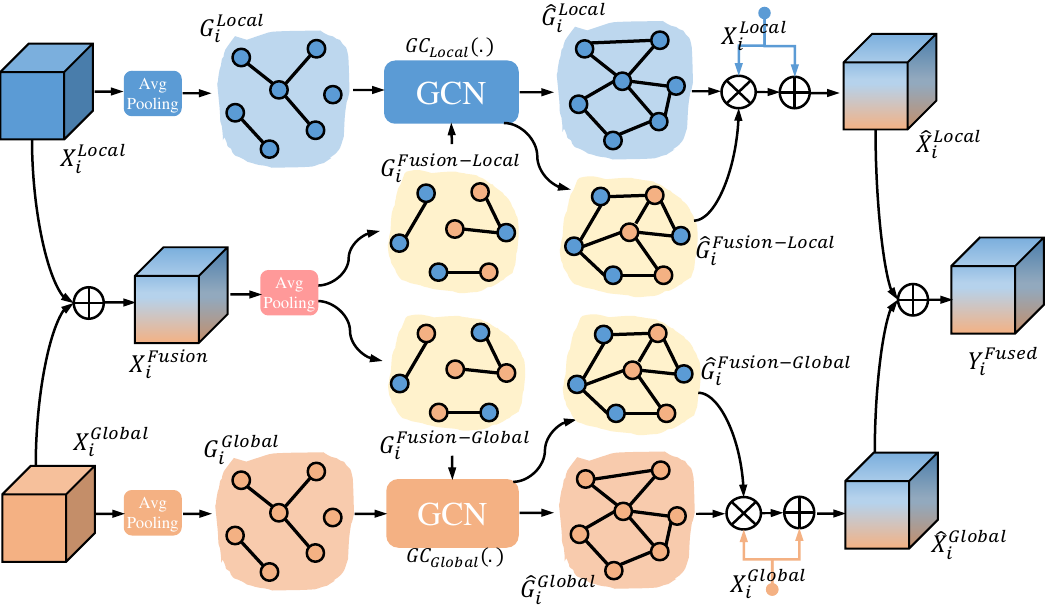}
	\caption{The architecture of the proposed Selective Graph Attention Fusion (SGAF) module.}
	\label{SGAF}
\end{figure}

In Fig. \ref{SGAF}, we have two types of input features: local features $X^{Local}_{i}$ obtained through plain CNN convolution, and global context features $X^{Global}_{i}$ obtained through capsule convolution. Then we add $X^{Local}_{i}$ and $X^{Global}_{i}$ to obtain the fused feature $X^{Fusion}_{i}$. We then apply three independent Average Pooling operators to eliminate spatial dimensions, preserving only the channel dimension. After pooling, we construct graphs along the channel dimension of the three features, resulting in four independent graphs: $G^{Local}_{i}$, $G^{Global}_{i}$, $G^{Fusion-Local}_{i}$ and $G^{Fusion-Global}_{i}$. The two graphs $G^{Fusion-Local}_{i}$ and $G^{Fusion-Global}_{i}$ constructed from $X^{Fusion}_{i}$ provide shared fusion information for both the local feature $X^{Local}_{i}$ and global feature $X^{Global}_{i}$. We assume that the two graphs should contain different topological structures of channels from $X^{Local}_{i}$ and $X^{Global}_{i}$ after learning and reweighting the graph convolution operators.
\begin{equation}
	\begin{aligned}
		&G_{i}^{\alpha} = AvgPooling(X_{i}^{\alpha}) \quad (\alpha \in [Global, Local])  \\
		&\widehat{G}^{\alpha}_{i} = GC_{\alpha}(G_{i}^{\alpha}) \quad (\alpha \in [Global, Local]) 
	\end{aligned}
\end{equation}
The graph represents each channel of the feature as a node. To learn the connectivity and relationships among nodes (channels), we apply only two graph convolution operators on the four constructed graphs: $GC_{Local}(.)$ for $G^{Local}_{i}$ and $G^{Fusion-Local}_{i}$, and $GC_{Global}(.)$ for $G^{Global}_{i}$ and $G^{Fusion-Global}_{i}$. By using shared graph convolution, the local or global graphs can share nodes and connectivity information with the fusion graphs, resulting in better connectivity weight adjustment, allowing more informative representation flow on the graph, and reducing computational cost and parameters.
\begin{equation}
	\begin{aligned}
		&G_{i}^{Fusion-\alpha} = Split(AvgPooling(X_{i}^{Fusion-\alpha})) \\
		&\widehat{G}^{Fusion-\alpha}_{i} = GC_{\alpha}(G_{i}^{Fusion-\alpha}) \quad (\alpha \in [Global, Local]) 
	\end{aligned}
\end{equation}
After applying graph convolution, we obtain four output graphs: $\widehat{G}^{Local}_{i}$, $\widehat{G}^{Fusion-Local}_{i}$, $\widehat{G}^{Global}_{i}$, and $\widehat{G}^{Fusion-Global}_{i}$. We then apply $\widehat{G}^{Local}_{i}$ and $\widehat{G}^{Global}_{i}$ on the input features $X^{Local}_{i}$ and $X^{Global}_{i}$ using multiplication and addition operators, respectively, which can be viewed as a kind of self-attention because the graph attention weights generated from the input features are applied back on the channels of original input features. At the same time, $\widehat{G}^{Fusion-Local}_{i}$ and $\widehat{G}^{Fusion-Global}_{i}$ are applied on the input features $X^{Local}_{i}$ and $X^{Global}_{i}$ using multiplication operators. The resulted refined output features are denoted as $\widehat{X}^{Local}_{i}$ and $\widehat{X}^{Global}_{i}$.
\begin{equation}
	\begin{aligned}
		&\widehat{X}^{\alpha}_{i} = X_{i}^{\alpha} * Expand(\widehat{G}_{i}^{\alpha} * \widehat{G}_{i}^{Fusion-\alpha}) + X_{i}^{\alpha} \\  & (\alpha \in [Global, Local])
	\end{aligned}
\end{equation}
Finally, we add $\widehat{X}^{Local}_{i}$ and $\widehat{X}^{Global}_{i}$ together to obtain the fused feature $Y^{Fused}_{i}$.
\begin{equation}
	\begin{aligned}
		&Y^{Fused}_{i} = \widehat{X}^{Local}_{i} + \widehat{X}^{Global}_{i}
	\end{aligned}
\end{equation}

\subsection{Bottleneck Graph Attention Module}

\begin{figure*}[htbp]
	\centering
	\includegraphics[width=13cm]{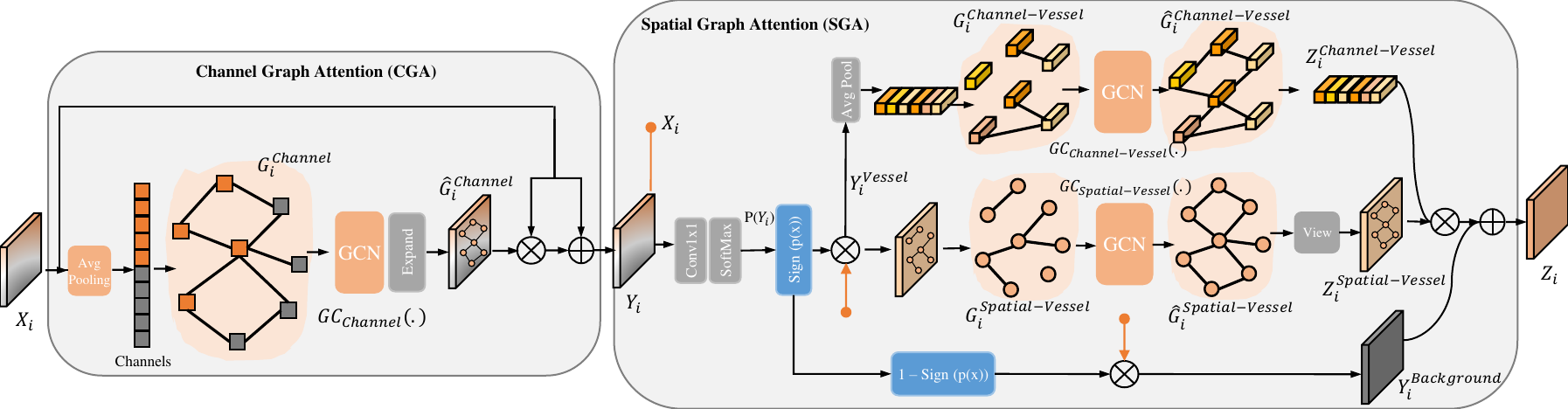}
	\caption{The architecture of the proposed Bottleneck Graph Attention (BGA) module.}
	\label{BGA}
\end{figure*}

To improve vessel continuity, particularly for thin vessels, we propose a novel Bottleneck Graph Attention (BGA) module comprising of Channel-wise Graph Attention (CGA) and Spatial Graph Attention (SGA). The input features $X_{i}$ are first fed into CGA, where an Average Pooling operator is used to extract channel-only features, transforming the feature shape from $[B, C, H, W]$ to $[B, C, 1, 1]$. A graph $G^{Channel}_{i}$ is constructed along the channel dimension, where each node represents a channel of features and the edge connectivity between nodes indicates their relationship. A graph convolution operator $GC_{Channel}(.)$ is applied to $G^{Channel}_{i}$, producing an output graph $\widehat{G}^{Channel}_{i}$ with re-weighted connectivity and re-modelled channel relationships. The refined graph $\widehat{G}^{Channel}_{i}$ is then expanded along the spatial dimensions and recovered to $[B, C, H, W]$. 
\begin{equation}
	\begin{aligned}
		&\widehat{G}^{Channel}_{i} = GC_{Channel}(AvgPooling(X_{i}))
	\end{aligned}
\end{equation}
The refined feature and graph representation $\widehat{G}^{Channel}_{i}$ are fused with the input feature $X{i}$ through multiplication and addition, generating the output feature $Y_{i}$. The CGA module enables the representation of channel dependencies as a graph and captures the relationships among channels.
\begin{equation}
	\begin{aligned}
		&Y_{i} = X_{i} * Expand(\widehat{G}^{Channel}_{i}) + X_{i}
	\end{aligned}
\end{equation}
In the SGA module, the input feature is $Y_{i}$, and a feature selector is proposed to extract vessels from the fundus background. The feature selector applies a conv1x1 $Conv(.)$ operator to reduce the dimension of $Y_{i}$ and Softmax function $Softmax(.)$ to calculate a probability map $p(Y_{i})$, which contains information about the probability that each pixel belongs to a vessel, ranging from 0 to 1.
\begin{equation}
	p(Y_{i}) = \textit{Softmax}(Conv(Y_{i}))
\end{equation}
A pre-defined piecewise function called the Sign function 
called the Sign function $Sign(.)$ is then applied to separate the probability values into two intervals. Specifically, a threshold of 0.4 was set in our experiments, indicating that the pixels with probability values greater than 0.4 correspond to blood vessel pixels, while those with values less than 0.4 correspond to background pixels. This allows for effective separation of vessel regions from the background.
The $Sign(.)$ function is defined as
\begin{equation}
	Sign(x)=
	\begin{cases}
		1 \quad \quad {\quad x > 0.4 \quad \quad (Vessel)}\\
		0 \quad \quad {\quad x < 0.4 \quad \quad (Background)}\\
	\end{cases}
\end{equation}
where $x$ means the probability of each pixel in the probability map $p(Y_{i})$. 
Using the Sign function, we can obtain vessel features $Y^{Vessel}_{i}$ and background features $Y^{Background}_{i}$ separately from input features $Y_{i}$ based on their probability values. 
\begin{equation}
	Y^{Vessel}_{i}, \quad Y^{Background}_{i} = \textit{Sign}(Y_{i})
\end{equation}
To improve the continuity of vessels, we perform two individual operations. The first operation involves constructing a graph $G^{Spatial-Vessel}_{i}$ for the vessel-only features based on their spatial distribution. Nodes and edges in the graph represent the vessels and their connectivity, respectively. We then apply a graph convolution $GC_{Spatial-Vessel}(.)$ on the graph $G^{Spatial-Vessel}_{i}$ to learn information about the nodes and edges connectivity, aiming to improve the continuity of vessels without interference from the background, especially noise and other tissues in the background. This yields the output features of vessels $Z^{Spatial-Vessel}_{i}$.
\begin{equation}
	Z^{Spatial-Vessel}_{i} = GC_{Spatial-Vessel}(G^{Spatial-Vessel}_{i})
\end{equation}
In addition to improving vascular continuity in spatial distribution, we also enhance semantic consistency. To achieve this, we use an average pooling operator to extract channel information of vessels, and construct a graph $G^{Channel-Vessel}_{i}$ for these channels. We then apply a graph convolution operator $GC_{Channel-Vessel}(.)$ to learn the graph representation of channels, yielding the output features of vessels $Z^{Channel-Vessel}_{i}$.
\begin{equation}
	Z^{Channel-Vessel}_{i} = GC_{Channel-Vessel}(G^{Channel-Vessel}_{i})
\end{equation}
Finally, we multiply $Z^{Spatial-Vessel}_{i}$ and $Z^{Channel-Vessel}_{i}$, add $Y^{Background}_{j}$, and then obtain the refined features $Z_{i}$ whose vascular continuity has been enhanced.
\begin{equation}
	Z_{i} = Z^{Channel-Vessel}_{i} * Z^{Spatial-Vessel}_{i} + Y^{Background}_{i}
\end{equation}
Through this approach, we can improve the connectivity of vessels without being affected by other tissues.

\subsection{Multi-Scale Graph Fusion Module}

\begin{figure}[htbp]
	\centering
	\includegraphics[width=12cm]{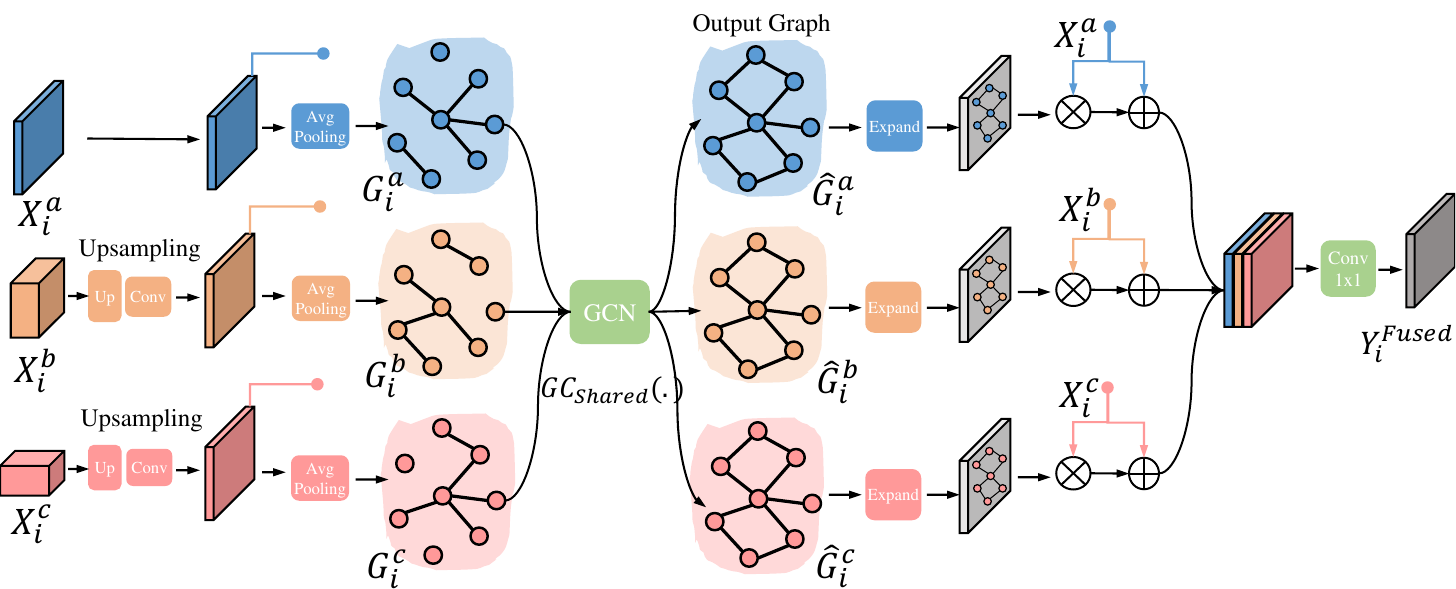}
	\caption{The proposed Multi-Scale Graph Fusion (MSGF) module.}
	\label{MSGF}
\end{figure}

To integrate the multi-scale features extracted from different stages of the UNet, we propose a Multi-Scale Graph Fusion module, as shown in Fig. \ref{MSGF}. The input features $X^{a}_{i}$, $X^{b}_{i}$ and $X^{c}_{i}$ are obtained from different upsampling stages of the UNet. Firstly, we apply upsampling and conv1x1 operators on $X^{b}_{i}$ and $X^{c}_{i}$ to reshape their spatial and channel dimensions to match those of $X^{a}_{i}$. 
\begin{equation}
	\begin{aligned}
		&G_{i}^{\alpha} = AvgPooling(Up(Conv1x1(X_{i}^{\alpha}))) \quad (\alpha \in [b, c])
	\end{aligned}
\end{equation}
Subsequently, we apply Average Pooling operators on these features to reduce their dimensions and preserve only channel-wise information. Then, we construct three independent graphs, $G^{a}_{i}$, $G^{b}_{i}$ and $G^{c}_{i}$, for these features along the channel-wise dimension.

Instead of adopting three individual graph convolution on these three independent graphs, we use only a single shared graph convolution $GC_{Shared}(.)$ to conduct convolutional process on $G^{a}_{i}$, $G^{b}_{i}$ and $G^{c}_{i}$, because we assume that graphs constructed from different scales with the same input feature have similar graph patterns and node connectivities. 
Adopting shared graph convolution can simultaneously capture the topological structure representations of $G^{a}_{i}$, $G^{b}_{i}$ and $G^{c}_{i}$, and adjust the connectivity of nodes on the graph by taking other graphs into the consideration, so that the information can propogate and flow on the graphs constructed from multi-scale features. 
After applying graph convolution operators, we obtain three independent graph representations $\widehat{G}^{a}_{i}$, $\widehat{G}^{b}_{i}$ and $\widehat{G}^{c}_{i}$. 
\begin{equation}
	\begin{aligned}
		&\widehat{G}_{i}^{\alpha} = GC_{Shared}(G_{i}^{\alpha}) \quad (\alpha \in [a, b, c])
	\end{aligned}
\end{equation}
And then these ouput channel-wise graphs are expanded spatially and applied directly on each input feature $X^{a}_{i}$, $X^{b}_{i}$ and $X^{c}_{i}$, obtaining three refined features $\widehat{X}^{a}_{i}$, $\widehat{X}^{b}_{i}$ and $\widehat{X}^{c}_{i}$, respectively. 
\begin{equation}
	\begin{aligned}
		&\widehat{X}^{\alpha}_{i} = X_{i}^{\alpha} * Expand(\widehat{G}_{i}^{\alpha}) + X_{i}^{\alpha} \quad (\alpha \in [a, b, c])
	\end{aligned}
\end{equation}
Finally, we concatenate the three refined features and adopt a conv1x1 operator to reduce the dimension and generate the output fused feature $Y^{Fused}_{i}$.
\begin{equation}
	\begin{aligned}
		&Y^{Fused}_{i} = Conv1x1(Concat(\widehat{X}^{\alpha}_{i})) \quad (\alpha \in [a, b, c])
	\end{aligned}
\end{equation}

\subsection{Loss Function}
The Cross Entropy (CE) loss $\mathcal L_{CE}$ is used as the loss function of our GCC-UNet, which is defined as
\begin{equation}
	\mathcal L_{CE}(p,q) =  -\sum_{k=1}^{N}p_{k}\ast log(q_{k})
\end{equation}

\section{Datasets and Materials}
\label{sec:datasets}

\subsection{Retinal Fundus Datasets}
Our GCC-UNet model was evaluated on three publicly available retinal vessel datasets, namely DRIVE \cite{staal2004ridge}, STARE \cite{hoover2000locating}, and CHASEDB1 \cite{fraz2012ensemble}.

The DRIVE dataset consists of 40 pairs of fundus images with a unified size of 565 $\times$ 584 pixels, where 20 pairs are used as training data and the remaining pairs as the test data. The STARE dataset comprises 20 pairs of fundus images and their corresponding labels with a size of 700 $\times$ 605 pixels. The first 10 pairs are used as the training dataset, and the remaining pairs are used as the test dataset. The CHASEDB1 dataset contains 28 pairs of fundus scans and their labels with a resolution of 999 $\times$ 960 pixels, where the first 20 pairs are used as the training dataset and the remaining 8 pairs are used as the test set.

Furthermore, to comprehensively evaluate our proposed GCC-UNet model, we also tested it on some challenging datasets, including AV-WIDE \cite{estrada2015retinal}, UoA-DR \cite{chalakkal2017comparative}, and UK Biobank \cite{sudlow2015uk}. It should be noted that the model used for testing on these datasets was trained on the DRIVE dataset.

\subsection{Evaluation Metrics}
We evaluated our model with some classical metrics, including F1 score (F1), accuracy (Acc), sensitivity (SE), specificity (SP), and area under the ROC curve (AUROC), which are defined as follows:
\begin{equation}
	\begin{aligned}
	&SE = Rec = \frac{TP}{TP + FN} \qquad SP = \frac{TN}{TN+FP} \\ \\
	&F1 = 2 \times \frac{Pre\times Rec}{Pre+Rec}\quad Acc = \frac{TP+TN}{TP+TN+FP+FN} 
		\end{aligned}
\end{equation}
where $TP$, $TN$, $FP$, and $FN$ represent the number of true positive,
true negative, false positive, and false negative pixels, respectively. $Pre$ and $Rec$ mean the precision and recall metrics, respectively. 
In addition, some advanced metrics proposed by Gegundez et al. \cite{gegundez2011function} were also adopted to evaluate our model, including connectivity (C), overlapping area (A), consistency of vessel length (L), and the overall metric (F). The overall metric (F) is defined as
\begin{equation}
	F = C \times A \times L
\end{equation}
Furthermore, rSE, rSP and rAcc proposed in \cite{yan2017skeletal} were also adopted to act as indicators for the evaluation, as well as the Matthews Correlation Coefficient (Mcc) \cite{khan2020hybrid}.

\section{Experiments}
\label{sec:experiments}

\subsection{Implementation details}
The GCC-UNet model was implemented using the PyTorch framework and trained on a TITAN XP GPU. During the training process, we used the Adam optimizer. The model was trained with a batch size of 32 over a total of 60 epochs. The early stopping strategy was adopted with a patience of 10 epochs. When calculating performance metrics, we only take pixels in the field of view (FOV) into consideration.

\subsection{Overall comparison with other methods}
\begin{figure}[htbp]
	\centering
	\includegraphics[width=13cm]{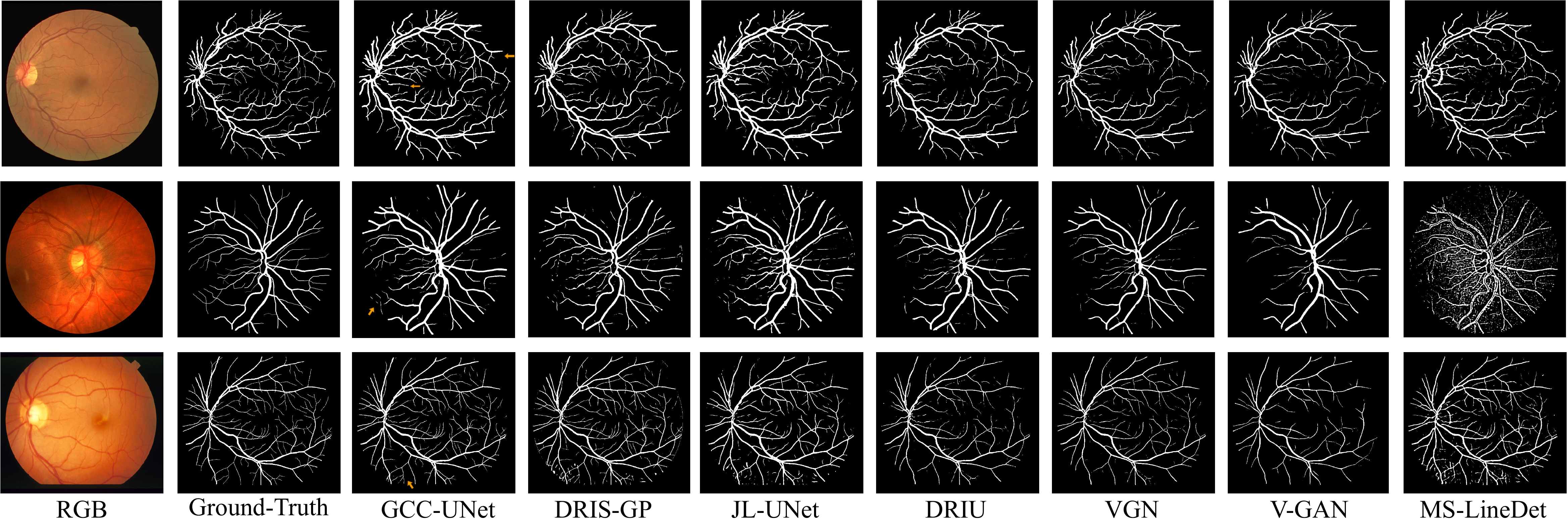}
	\caption{Visual comparison with other state-of-the-art methods on DRIVE(the first row), CHASEDB1(the second row) and STARE(the third row) datasets.}
	\label{ex1}
\end{figure}

\begin{figure}[htbp]
	\centering
	\includegraphics[width=13cm]{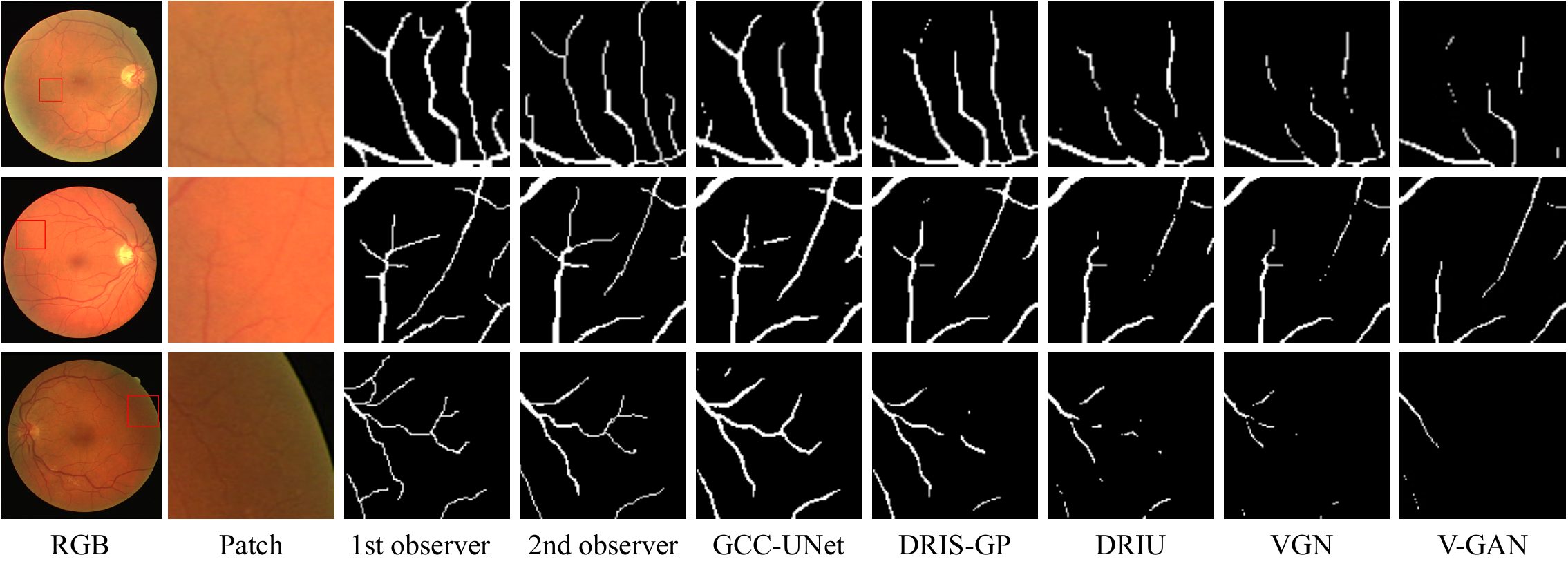}
	\caption{Visual comparison with other methods in terms of thin vessels.}
	\label{ex2}
\end{figure}

We conducted comprehensive comparison experiments to demonstrate the performance of our proposed GCC-UNet model. The results presented in Tables \ref{drive}, \ref{stare}, \ref{chasedb}, and \ref{drive2} show that our method outperforms numerous state-of-the-art methods on the DRIVE \cite{staal2004ridge}, STARE \cite{hoover2000locating}, and CHASEDB1 \cite{fraz2012ensemble} datasets, in terms of both traditional and advanced metrics. Furthermore, as shown in Fig. \ref{ex1} and \ref{ex2}, our method also exhibits superior visual performance compared with other methods, particularly in detecting thin vessels. These results provide further evidence of the effectiveness of our approach and its ability to capture global context and improve the continuity of vessels.

\begin{table}[htbp]
	\footnotesize
	\renewcommand\arraystretch{0.8}
	\setlength{\tabcolsep}{5pt}
	\centering
	\caption{Quantitative evaluation against other leading methods on \textbf{DRIVE}. \color{red}Red: the best, \color{blue}Blue: the second best.}
	\begin{tabular}{c|ccccc}
		\toprule
		Method       & F1  & Se & Sp & Acc & AUROC \\ \midrule

		2nd observer \cite{staal2004ridge}   &N.A  & 77.60 & 97.24 & 94.72 & N.A   \\

		HED \cite{xie2015holistically}  & 80.89 & 76.27 & 98.01 & 95.24 & 97.58 \\

		DeepVessel \cite{fu2016deepvessel}  & N.A & 76.12 & 97.68 & 95.23 & 97.52  \\

		Orlando et al. \cite{orlando2016discriminatively}    &N.A    &78.97    &96.84     &94.54     &95.06  \\

		JL-UNet \cite{yan2018joint}   &N.A    &76.53    &98.18     &95.42     &97.52   \\

		CC-Net \cite{feng2020ccnet}      &N.A    &76.25    &98.09     &95.28     &96.78   \\

		Att UNet \cite{oktay2018attention}                   
		&82.32  &79.46    &97.89    &95.64     &97.99   \\

		
		Yan et al. \cite{yan2018three}     &N.A     &76.31    &98.20    &95.33     &97.50   \\

		BTS-DSN \cite{guo2019bts}    &82.08     &78.00    &98.06    &95.51     &97.96    \\ 
		
		
		CTF-Net \cite{wang2020ctf}    &82.41    &78.49    &98.13     &\color{blue}95.67     &97.88   \\ 
		
		CSU-Net \cite{wang2020csu}         &\color{blue}82.51    &\color{red}80.71    &97.82     &95.65     &\color{blue}98.01 \\ 

  		RCAR-UNet \cite{ding2024rcar}    &80.47     &74.87    &\color{red}98.36    &95.37     &N.A    
  
  \\ \midrule
		\textbf{GCC-UNet (Ours)}  &\color{red}82.78   &\color{blue}80.32   &\color{blue}98.21  &\color{red}95.74  &\color{red}98.13 \\  \bottomrule

	\end{tabular}
	\label{drive}
\end{table}

\begin{table}[htbp]
	\footnotesize
	\renewcommand\arraystretch{0.8}
	\setlength{\tabcolsep}{5pt}
	\centering
	\caption{Quantitative comparison with other methods on \textbf{STARE}.}
	\begin{tabular}{c|ccccc}
		\toprule
		Method   & F1  & Se & Sp & Acc & AUROC \\ \midrule


		HED \cite{xie2015holistically} & \color{blue}82.68 &\color{red} 80.76 & 98.22 & 96.41 & 98.24 \\


		Orlando et al. \cite{orlando2016discriminatively}  &N.A    &76.80    &97.38     &95.19     &95.70  \\

		JL-UNet \cite{yan2018joint}     &N.A    &75.81    &98.46     &96.12     &98.01   \\

		Att UNet \cite{oktay2018attention} 
		&81.36     &\color{blue}80.67    &98.16    &96.32     &98.33   \\

		CC-Net \cite{feng2020ccnet}            
		&N.A     &77.09    &98.48    &96.33     &97.00   \\

		Dense UNet \cite{li2018h}                          
		&82.32     &78.59    &98.42    &\color{blue}96.44     &\color{blue}98.47   \\ 
		
		Yan et al. \cite{yan2018three}       &N.A     &77.35    &98.57    &96.38     &98.33   \\

		
		DUNet \cite{jin2019dunet}           &82.30     &78.92    &98.16    &96.34     &98.43   \\  

  RCAR-UNet \cite{ding2024rcar}       &78.50     &69.79    &\color{red}99.05    &95.94     &N.A   \\
		\midrule
		
		\textbf{GCC-UNet (Ours)}  &\color{red}82.82   &78.06   &\color{blue}98.77 &\color{red}96.58  &\color{red}98.56    \\
		
		\bottomrule

	\end{tabular}
	\label{stare}
\end{table}

\begin{table}[htbp]
	\footnotesize
	\renewcommand\arraystretch{0.8}
	\setlength{\tabcolsep}{10pt}
	\centering
	\caption{Quantitative comparison with other methods on \textbf{CHASEDB1}.}
	\begin{tabular}{c|ccccc}
		\toprule
		Method   & F1  & Se & Sp & Acc & AUROC \\ \midrule

		2nd observer \cite{staal2004ridge}  &N.A  & \color{blue}81.05 & 97.11 & 95.45 & N.A   \\

		HED \cite{xie2015holistically}  & 78.15 & 75.16 & 98.05 & 95.97 & 97.96 \\

		DeepVessel \cite{fu2016deepvessel}  & N.A & 74.12 & 97.01 & 96.09 & 97.90  \\

		Orlando et al. \cite{orlando2016discriminatively}    &N.A     &75.65     &96.55      &94.67    &94.78  \\

		JL-UNet \cite{yan2018joint} &N.A     &76.33     &\color{blue}98.09      &96.10     &97.81   \\

		Att UNet \cite{oktay2018attention}                
		&\color{blue}80.12     &80.10     &98.04     &96.42      &98.40   \\

		Dense UNet \cite{li2018h}                 
		&79.01     &78.93     &97.92     &96.11      &98.35   \\ 
		
		Yan et al. \cite{yan2018three} &N.A     &76.41     &98.06     &96.07      &97.76   \\

		BTS-DSN \cite{guo2019bts}  &\color{blue}79.83    &78.88   &98.01   &\color{blue}96.27 &\color{blue}98.40    \\ 
		
		DUNet \cite{jin2019dunet}         &79.32     &77.35     &98.01     &96.18      &98.39   \\ 

    RCAR-UNet \cite{ding2024rcar}       &74.70     &74.75    &97.98    &95.66     &N.A   \\

		\midrule
		
		
		\textbf{GCC-UNet (Ours)}  &\color{red}80.86  &\color{red}81.23   &\color{red}98.15   &\color{red}96.59    &\color{red}98.50    \\  \bottomrule

	\end{tabular}
	\label{chasedb}
\end{table}

\begin{table}[]
	\footnotesize
	\renewcommand\arraystretch{0.8}
	\setlength\tabcolsep{5pt}
	\centering
	\caption{Quantitative comparison with other methods in terms of metrics in \cite{yan2017skeletal} on \textbf{DRIVE} dataset.}
	\begin{tabular}{c|cccccccc}
		\toprule
		Method               & F    & C    & A    & L   & rSe  &rSp  &rAcc  &Mcc \\
		\midrule
		2nd observer        & 83.75 & 100 & 93.98 & 89.06 & 85.84  &99.19   &95.74   &76.00 \\ \midrule
		
		HED \cite{xie2015holistically}       & 80.09 & 99.75 & 90.06 & 89.11 & 71.57 &95.11   &89.08   &66.00 \\
		DRIU \cite{maninis2016deep}       & 80.43 & 99.56 & 91.52 & 88.23 & \color{blue}82.36  &96.85   &93.13  &71.61 \\
		DeepVessel \cite{fu2016deepvessel}       & 61.74 & 99.60 & 84.23 & 73.38 & 54.93  &99.78   &88.32   &73.34 \\
		V-GAN \cite{son2017retinal}       & 84.82 & 99.64 & 94.69 & \color{blue}89.84 & 80.77  &\color{blue}99.63   &94.76   &80.24 \\
		
		JL-UNet \cite{yan2018joint}       & 81.06 & 99.61 & 93.08 & 87.35 & 76.11  &99.57   &93.53   &78.98 \\
		SWT-FCN \cite{oliveira2018retinal}        &83.92       &\color{blue}99.73       &94.36       &89.11       &79.63     &99.64   &94.48   &\color{blue}80.53    \\
		DeepDyn \cite{khanal2020dynamic}          &84.53       &90.70       &94.58       &89.61       &81.52   &99.44  &\color{blue}94.82  &80.02       \\
		DAP \cite{sun2021robust}      &82.55       &99.72       &93.74       &88.24       &78.57   &99.57    &94.15    &79.00       \\
		DRIS-GP \cite{cherukuri2019deep}          &\color{blue}84.94      &99.68       &\color{blue}94.91       &89.74       &80.22  &\color{red}99.64   &94.66   &\color{red}81.84     \\
		\midrule
		\textbf{GCC-UNet}  &\color{red}85.83       &\color{red}99.75       &\color{red}95.10       &\color{red}90.46       &\color{red}82.60  &99.17   &\color{red}95.06  &80.27  \\  \bottomrule
		
	\end{tabular}
	\label{drive2}
\end{table}

\subsection{Comparison and ablation study of individual module}

\subsubsection{Comparison and ablation analysis between the proposed GC Conv and plain Capsule Conv}
To advance beyond the limitations of vanilla capsule convolution \cite{sabour2017dynamic}, we introduce the Graph Capsule Convolution (GC Conv), designed to capture the intricate interdependencies among channels, capsules, and even atomic units. In our experiments, we substituted the conventional convolution operations with both capsule convolution (Cap Conv) \cite{sabour2017dynamic} and our innovative GC Conv within the U-Net framework. As illustrated in Table \ref{tb_gc_conv}, the performance notably declined when vanilla convolutions were replaced with Cap Conv, whereas our GC Conv demonstrated substantial improvements.

This enhancement is attributed to the fact that while capsule convolution primarily focuses on capturing global features such as relative positions, orientations, and colors of vessels, it does not explicitly model the interactions among these global attributes. For example, capillaries typically exhibit lighter colors and are found at terminal branches (positions), with more intricate orientations. In contrast, GC Conv excels by modeling the relationships among these characteristics and learning their correlations in a graph-based framework, thereby capturing more comprehensive and nuanced feature interdependencies.

\begin{table}[]
	\footnotesize
	\renewcommand\arraystretch{1.1}
	\setlength{\tabcolsep}{5pt}
	\centering
	\caption{Comaprison between the vanilla Capsule Conv (Cap Conv) in \cite{sabour2017dynamic} and our Graph Capsule Conv (GC Conv) on DRIVE.}
	\begin{tabular}{c|ccccc}
		\toprule
		Method                     & F1    & Se    & Sp    & Acc   & AUROC   \\
		\midrule
		Baseline (UNet) \cite{ronneberger2015u}            & \color{blue}81.76 & \color{red}78.36 & 98.03 & \color{blue}95.56 & \color{blue}97.86 \\
		+ Capsule Conv \cite{sabour2017dynamic}  &81.19      &78.07      &\color{blue}98.12       &95.53       &97.81       \\
		\midrule
		+ \textbf{GC Conv (Proposed)} &\color{red}82.01       &\color{blue}78.12       &\color{red}98.18       &\color{red}95.63       &\color{red}97.93    \\ 
		\bottomrule
	\end{tabular}
	\label{tb_gc_conv}
\end{table}

\subsubsection{Comparison and ablation analysis between the proposed SGAF and other fusion modules}
We conducted an extensive series of experiments to assess the efficacy of our proposed Selective Graph Attention Fusion (SGAF) module in comparison to other fusion strategies. Table \ref{tb_sgaf} presents the performance metrics for integrating local features with various types of global features. These global features were extracted using the conventional Capsule Convolution (Cap Conv) with dynamic routing \cite{sabour2017dynamic} and our novel Graph Capsule Convolution (GC Conv) with graph-based dynamic routing. Alongside SGAF, we also assessed the performance of vanilla Conv1x1 \cite{simonyan2014very} and Selective Kernel Attention (SK) \cite{li2019selective} as alternative fusion modules.

The results in Table \ref{tb_sgaf} indicate that Conv1x1 was suboptimal for fusing local and global features, failing to distinguish between beneficial channels in these feature types. Conversely, SK Attention demonstrated effective feature fusion across different mechanisms, achieving commendable performance. Nonetheless, our SGAF module surpassed SK Attention by a substantial margin. Additionally, GC Conv significantly outperformed Cap Conv in extracting global contextual features while utilizing the same fusion module.

\begin{figure}[htbp]
	\centering
	\includegraphics[width=10cm]{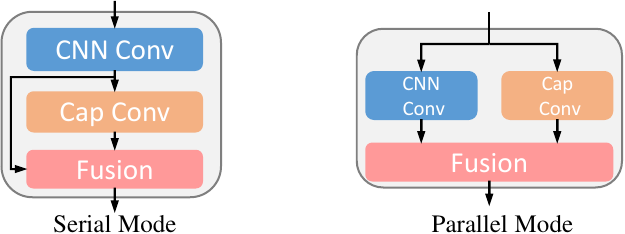}
	\caption{Different modes for combining CNN Conv and Capsule Conv.}
	\label{mode}
\end{figure}

We also investigated various operational modes, including the serial and parallel configurations depicted in Fig. \ref{mode}, for combining CNN Conv and Capsule Conv. Our findings reveal that the serial mode outperforms the parallel mode. Given that both Cap Conv and GC Conv cannot directly extract global information from raw images, the most effective strategy involves initially using vanilla CNN convolutions to extract features, followed by capsule convolutions to further refine global contextual information from the CNN features. This approach is then complemented by fusing local and global features through skip connections.

\begin{table}[htbp]
	\footnotesize
	\renewcommand\arraystretch{1.1}
	\setlength{\tabcolsep}{4pt}
	\centering
	\caption{Comaprison between our SGAF and other fusion modules (Conv1x1\cite{simonyan2014very}, SK Attention\cite{li2019selective}) on DRIVE.}
	\begin{tabular}{c|ccccc}
		\toprule
		Method                     & F1    & Se    & Sp    & Acc   & AUROC   \\
		\midrule
		Baseline (UNet) \cite{ronneberger2015u}            &81.76 & 78.36 & 98.03 & 95.56 & 97.86 \\
		Baseline (Capsule UNet) \cite{sabour2017dynamic}           &81.19   &78.07   &\color{blue}98.12   &95.53   &97.81 \\
		\midrule
		CAPSULE / FUSION                     & F1    & Se    & Sp    & Acc   & AUC   \\
		\midrule
		Cap Conv\cite{sabour2017dynamic} / Conv1x1\cite{simonyan2014very}     &81.42       &77.96       &97.79       &95.52       &97.82       \\
		Cap Conv\cite{sabour2017dynamic} / SK\cite{li2019selective}          &82.12       &78.16       &97.76       &95.65       &98.01       \\
		Cap Conv\cite{sabour2017dynamic} / \underline{\textbf{SGAF}}  &\color{blue}82.30       &\color{blue}78.98       &\color{red}98.18       &95.67       &\color{blue}98.04       \\
		\midrule
		\underline{\textbf{GC Conv}} / Conv1x1\cite{simonyan2014very}     &81.82       &78.53       &97.92       &95.59       &97.87       \\
		\underline{\textbf{GC Conv}} / SK\cite{li2019selective}         &82.23       &78.86       &97.91       &\color{blue}95.68       &98.03    \\
		\midrule
		\underline{\textbf{GC Conv}} / \underline{\textbf{SGAF}} \textbf{(Parallel)}  &81.75       &\color{blue}79.36       &98.05       &95.64       &97.99      \\
		\underline{\textbf{GC Conv}} / \underline{\textbf{SGAF}} \textbf{(Serial)}  &\color{red}82.42       &\color{red}79.45       &98.11       &\color{red}95.70       &\color{red}98.07       \\
		\bottomrule
	\end{tabular}
	\label{tb_sgaf}
\end{table}

\subsubsection{Comparison and ablation analysis between our proposed BGA and other attention modules in the bottleneck}
We conducted a series of experiments to evaluate the performance of our proposed Bottleneck Graph Attention (BGA) module in comparison with several prominent attention mechanisms. As demonstrated in Table \ref{tb_bga}, our BGA module consistently outperforms other well-established attention modules, including SE \cite{hu2018squeeze}, CBAM \cite{woo2018cbam}, Non-Local \cite{wang2018non}, and Self-Attention \cite{vaswani2017attention}. Additionally, both our Channel Graph Attention (CGA) and Spatial Attention (SGA) components achieve notable performance.

The superior performance of our BGA module is attributed to its ability to model channel relationships through CGA by constructing and learning a graph representation. Furthermore, BGA leverages SGA to effectively distinguish vessels from the background and enhance vessel continuity by learning the connectivity among vessel nodes. This dual approach enables more accurate vessel segmentation and continuity preservation, highlighting the efficacy of our proposed attention mechanism.

\begin{table}[htbp]
	\footnotesize
	\renewcommand\arraystretch{1.1}
	\centering
	\caption{Comaprison and ablation study of the proposed BGA and other attention modules on DRIVE.}
	\begin{tabular}{c|ccccc}
		\toprule
		Method                     & F1    & Se    & Sp    & Acc   & AUROC   \\
		\midrule
		Baseline (UNet) \cite{ronneberger2015u}            &81.76 & 78.36 & \color{blue}98.03 & 95.56 & 97.86 \\
		\midrule
		+ SE \cite{hu2018squeeze}      &81.81       &79.03      &97.77       &95.60       &97.90      \\
		+ CBAM \cite{woo2018cbam}            &81.06       &78.85       &97.87       &95.61       &97.89       \\
		+ Non-Local \cite{wang2018non}       &81.75       &78.98       &97.76       &95.61       &97.91       \\
		+ Self-Attention \cite{vaswani2017attention}    &82.03   &79.45   &97.96   &95.64   &97.93       \\
		\midrule
		+ CGA (Proposed) &82.18       &79.32       &98.00       &95.64       &97.93       \\
		+ SGA (Proposed)  &\color{blue}82.11   &\color{red}79.89   &97.95  &\color{blue}95.64 &\color{blue}97.93     \\
		\midrule
		\textbf{+ BGA (Proposed)} &\color{red}82.25       &\color{blue}79.65      &\color{red}98.05       &\color{red}95.67       &\color{red}97.94  \\
		\bottomrule
	\end{tabular}
	\label{tb_bga}
\end{table}

\subsubsection{Comparison and ablation analysis between our proposed MSGF and other multi-scale fusion modules}
We conducted a series of experiments to assess the efficacy of our proposed Multi-Scale Graph Fusion (MSGF) module against other multi-scale fusion techniques, including the vanilla Conv1x1 and the fusion module described in \cite{wu2021region}. Furthermore, we evaluated three distinct modes of our MSGF module: Individual (applying separate graph convolutions to each of the three different scales of input graphs), Concat (concatenating the graphs from the three input features and processing them with a single graph convolution), and Shared (feeding the graphs from the input features into a shared graph convolution).

As illustrated in Table \ref{tb_msgf}, all three MSGF modes outperformed the other fusion modules. Among these, the Shared mode demonstrated superior performance with fewer parameters and reduced computational costs. This advantage arises because the Shared mode processes the graphs from the three scales using a single graph convolution, enabling the convolution operator to assimilate all relevant information and features concurrently. Additionally, since the features at different scales are derived from the same fundus image and are presumed to follow a similar graph pattern, utilizing a single graph convolution aligns the graph representations across these scales. This approach facilitates the integration and complementarity of information from all scales within one shared graph convolution, thereby enhancing overall performance.

\begin{table}[htbp]
	\footnotesize
	\renewcommand\arraystretch{1.1}
	\setlength{\tabcolsep}{5pt}
	\centering
	\caption{Ablation study of the proposed MSGF on DRIVE.}
	\begin{tabular}{c|ccccc}
		\toprule
		Method                     & F1    & Se    & Sp    & Acc   & AUROC   \\
		\midrule
		Baseline (UNet) \cite{ronneberger2015u}            &81.76 & 78.36 & \color{blue}98.03 & 95.56 & 97.86 \\
		\midrule
		+ Fusion via Conv1x1\cite{simonyan2014very}      &81.69     &78.88    &97.89    &96.59    &97.89    \\
		+ Fusion module in \cite{wu2021region}         &81.86    &78.54    &97.95    &96.62    &97.91    \\
		\midrule
		+ MSGF (Individual)  &\color{blue}82.03       &78.84       &98.02       &96.64       &97.93       \\
		+ MSGF (Concat)  &81.95     &\color{blue}79.14     &97.98       &\color{blue}96.64       &\color{blue}97.93       \\
		\midrule
		\textbf{+ MSGF (Shared)} &\color{red}82.15     &\color{red}79.23     &\color{red}98.08     &\color{red}95.68    &\color{red}97.94  \\
		\bottomrule
	\end{tabular}
	\label{tb_msgf}
\end{table}

\subsection{Overall ablation study of different fashions}

\begin{table}[]
	\footnotesize
	\renewcommand\arraystretch{0.8}
	\setlength{\tabcolsep}{3pt}
	\centering
	\caption{Overall ablation study of each proposed module on DRIVE.}
	\begin{tabular}{c|ccccc}
		\toprule
		Method (Local-only)                    & F1    & Se    & Sp    & Acc   & AUROC   \\
		\midrule
		Local UNet (Plain Conv) \cite{ronneberger2015u}            & 81.76 & 78.36 & 98.03 & 95.56 & 97.86 \\

		+ BGA &82.25       &79.65      &98.05       &95.67       &97.94  \\
		+ BGA + MSGF &82.42       &80.04      &98.09       &95.70       &98.01  \\
		\midrule
		\midrule
		Method (Vanilla Global-only)                    & F1    & Se    & Sp    & Acc   & AUC   \\
		\midrule
		Global UNet (Capsule Conv) \cite{sabour2017dynamic}        &81.19   &78.07   &98.12   &95.53    &97.81 \\

		+ BGA &81.43       &78.85      &98.05       &95.58       &97.87  \\
		+ BGA + MSGF &81.93       &79.32      &98.12       &95.63       &97.93  \\
		\midrule
		\midrule
			Method (Improved Global-only)                    & F1    & Se    & Sp    & Acc   & AUC   \\
		\midrule
		Global UNet (GC Conv)     &82.01   &78.12   &98.18    &95.63     &97.93 \\
		
		+ BGA &82.25       &79.65      &98.05       &95.67       &97.98  \\
		+ BGA + MSGF &82.36       &80.04      &98.09       &95.70       &98.03  \\
		\midrule
		\midrule
		Method (Global-Local Fusion)                    & F1    & Se    & Sp    & Acc   & AUC   \\
		\midrule
		Fusion UNet (Plain Conv + GC Conv)    &82.42      &79.45       &98.11      &95.70    &98.07  \\

		+ BGA &82.61       &80.02      &98.16       &95.71       &98.10  \\
		+ BGA + MSGF   &82.78   &80.32   &98.21  &95.74  &98.13 \\
		\bottomrule
	\end{tabular}
	\label{tb_abla_all}
\end{table}

Following a comprehensive ablation analysis of each proposed module, we conducted an overall ablation study in three different configurations: local-only, global-only, and global-local fusion. As detailed in Table \ref{tb_abla_all}, we examined four configurations: local-only, vanilla global-only, improved global-only, and global-local fusion.

In the local-only configuration, we used the vanilla U-Net as the baseline, which comprises basic convolutional blocks capable of capturing only local features. When augmented with our proposed Bottleneck Graph Attention (BGA) and Multi-Scale Graph Fusion (MSGF) modules, the model achieved a notable performance improvement, attaining 98.01\% AUROC.

For the vanilla global-only configuration, we substituted the standard convolution layers in U-Net with vanilla capsule convolution \cite{sabour2017dynamic} to create a new U-Net model focused solely on capturing global features. The performance of this global-only U-Net was inferior to that of the local-only configuration. However, when enhanced with our BGA and MSGF modules, the model's performance saw a significant boost. Furthermore, replacing the vanilla capsule convolution with our Graph Capsule Convolution (GC Conv) led to even better performance, demonstrating that GC Conv surpasses the vanilla capsule convolution \cite{sabour2017dynamic} in modeling contextual features and that BGA and MSGF effectively complement our GC Conv.

Finally, experiments with the global-local fusion baseline revealed that this configuration outperforms both the local-only and global-only models. This result underscores the substantial benefit of integrating both local and global features for retinal vessel segmentation, affirming the effectiveness of our fusion approach.

\subsection{Comparison study on challenging test sets}
To evaluate the generalization capabilities of our GCC-UNet model, we performed experiments on several challenging datasets, including AV-WIDE \cite{estrada2015retinal}, UoA-DR \cite{chalakkal2017comparative}, and UK Biobank \cite{sudlow2015uk}. For comparison, all models were trained from scratch on the DRIVE dataset.

\begin{figure}[htbp]
	\centering
	\includegraphics[width=13cm]{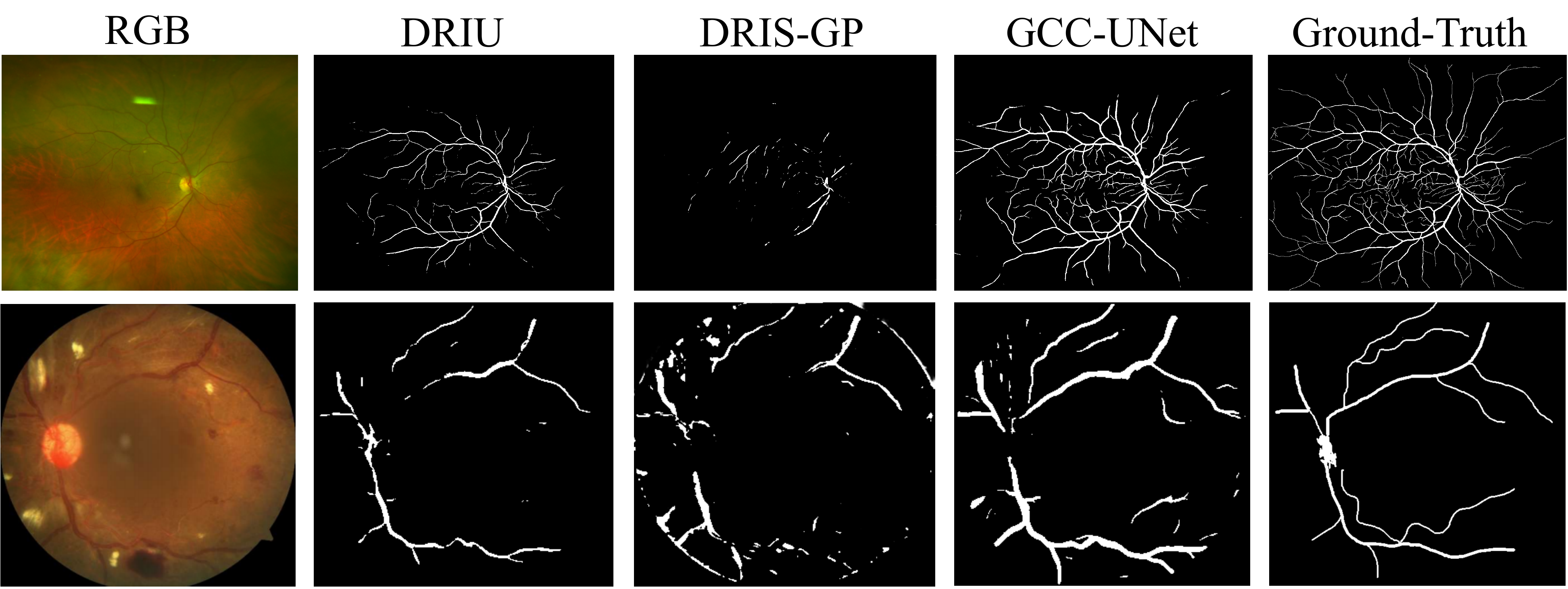}
	\caption{Visual comparison on the AV-WIDE (1st row) and UoA-DR (2nd row) datasets.}
	\label{avwide}
\end{figure}

\begin{figure}[htbp]
	\centering
	\includegraphics[width=13cm]{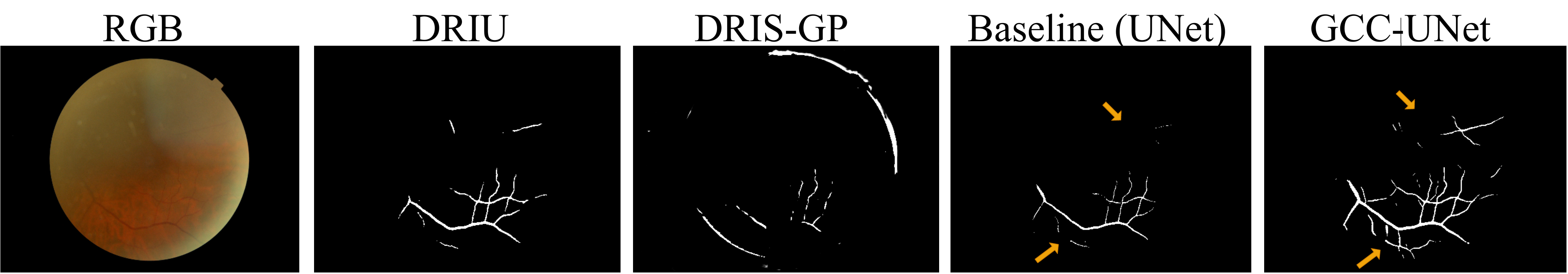}
	\caption{Visual comparison on the UK Biobank dataset.}
	\label{ukbb}
\end{figure}

Our results, as illustrated in Fig. \ref{avwide} and \ref{ukbb}, reveal that GCC-UNet surpasses state-of-the-art methods such as DRIU and DRIS-GP, and offers a substantial improvement over the baseline U-Net. In particular, the UK Biobank test results (Fig. \ref{ukbb}) highlight areas where thin vessels are obscured by opacities, which are notoriously difficult to detect even by human experts. Despite these challenges, our GCC-UNet model effectively identified these blurred and occluded vessels, demonstrating its superior performance and robustness.

\begin{table}[htbp]
	\footnotesize
	\renewcommand\arraystretch{1.1}
	\setlength{\tabcolsep}{4pt}
	\centering
	\caption{Comparison of Model Size, Parameters and Flops on DRIVE.}
	\begin{tabular}{c|ccccc}
		\toprule
		Method                     & UNet    & Att UNet    & Dense UNet    & DUNet   & GCC-UNet  \\
		\midrule
		Size (M)       & 3.4  & 7.1  & 11.0  &7.4    & 5.5 \\
		Params (M)       & 0.28  & 0.29  & 0.31  &0.43    & 0.39 \\
		Flops (G)            & 0.14   &0.15    & 0.44    &0.23  & 0.18  \\
		
		\bottomrule
	\end{tabular}
	\label{ex_para_flop}
\end{table}

\subsection{Comparison of model size, parameters and flops}

To highlight the efficiency of our GCC-UNet, we compared it with several UNet-based methods, including vanilla UNet \cite{ronneberger2015u}, Attention U-Net \cite{oktay2018attention}, Dense U-Net \cite{li2018h}, and Deformable U-Net \cite{jin2019dunet}, evaluating model size, parameter count, and computational complexity (FLOPs).

As presented in Table \ref{ex_para_flop}, GCC-UNet outperforms many existing UNet-based models while maintaining a compact parameter size and relatively small model footprint. This indicates that our GCC-UNet strikes an effective balance between computational efficiency and model performance, providing both high efficacy and manageable resource requirements.

%
%

\subsection{The potential of our method: Extend the ability of geometric modeling to boundary detection tasks}

We propose integrating graph-based and capsule-based approaches into medical image segmentation tasks, particularly those requiring precise boundary detection, such as optic disc segmentation, brain tumor segmentation, and biological cell segmentation. Our Bottleneck Graph Attention (BGA) module shows significant promise for enhancing boundary continuity. For instance, the sign function in the Spatial Graph Attention (SGA) can be employed to approximate boundary locations, followed by constructing a graph to reinforce continuity. Alternatively, the use of oriented kernels \cite{cherukuri2019deep} \cite{wei2023orientation} can further enhance boundary continuity.

Looking ahead, a promising direction for future work is to incorporate orientation modeling into graph construction processes. This enhancement could further improve the continuity of vessels and object boundaries, thereby advancing the accuracy and effectiveness of medical image segmentation tasks.

\section{Discussion}
\label{sec:discussion}
Our proposed GCC-UNet has achieved impressive results in retinal vessel segmentation by effectively integrating global context, part-to-whole relationships, and local-global fusion, while also enhancing vessel continuity. Notably, the model maintains a relatively compact parameter count of just 5.48M. However, there are inherent limitations in our approach. As highlighted by \cite{sabour2017dynamic}, although capsule convolution facilitates the capture of global context, it also substantially increases the model's computational cost, leading to slower inference speeds. This challenge is a fundamental characteristic of capsule convolution.

While our Graph Capsule Convolution (GC Conv) significantly enhances the efficacy of capsule convolution, it does not address the issue of increased computational cost or improved inference speed. Future work will focus on developing techniques to accelerate capsule convolution. Additionally, exploring the application of directed graph neural networks \cite{tong2020directed} in medical image segmentation tasks could be an exciting avenue for improving the continuity of curvilinear boundaries, further advancing the field.

\section{Conclusion}
\label{sec:conclusion}
In this study, we introduce a novel model for retinal vessel segmentation that combines global and local fusion within a U-Net framework, incorporating vanilla, graph, and capsule convolutions in a unified approach. This represents the first attempt to integrate these diverse convolutional techniques. Specifically, our model utilizes capsule convolution to capture global contextual information and graph convolution to model vessel connectivity and enhance continuity.
Our Graph Capsule Convolution (GC Conv) advances the traditional capsule convolution by improving its effectiveness. Additionally, the Selective Graph Attention Fusion (SGAF) module facilitates the integration of features across different domains (CNN, Graph, and Capsule). The Bottleneck Graph Attention (BGA) module enhances vessel continuity through a divide-and-conquer strategy, while the Multi-Scale Graph Fusion (MSGF) module effectively manages multi-scale feature fusion.
Crucially, the modules developed in this study are versatile and can be extended to a variety of applications beyond vessel segmentation. These include MRI tumor segmentation, geometric modeling of medical images, and both semantic and instance segmentation tasks.

\bibliographystyle{unsrt}
\bibliography{ref}

\end{document}